\documentclass[times,twocolumn,tighten]{aastex61}

\usepackage[modulo]{lineno}

\received{2017 June~16}
\revised{2018 January~8}
\accepted{2018 February~26}
\submitjournal{ApJS}

\shorttitle{Data Analysis and Management for High-Resolution Solar Physics}
\shortauthors{Denker et al.}

\begin{document}

%
%

\title{High-Cadence Imaging and Imaging Spectroscopy at the GREGOR Solar
    Telescope ---\\ A Collaborative Research Environment for High-Resolution 
    Solar Physics}

\correspondingauthor{Carsten Denker}
\email{cdenker@aip.de}

\author[0000-0002-7729-6415]{Carsten Denker}
\affil{Leibniz Institute for Astrophysics Potsdam (AIP),
    An der Sternwarte 16,
    14482 Potsdam, Germany}

\author[0000-0002-3242-1497]{Christoph Kuckein}
\affil{Leibniz Institute for Astrophysics Potsdam (AIP),
    An der Sternwarte 16,
    14482 Potsdam, Germany}

\author[0000-0003-1054-766X]{Meetu Verma}
\affil{Leibniz Institute for Astrophysics Potsdam (AIP),
    An der Sternwarte 16,
    14482 Potsdam, Germany}
    
\author[0000-0002-6546-5955]{Sergio J.\ Gonz\'alez Manrique}
\affil{Astronomical Institute of the Slovak Academy of Sciences, 
     05960 Tatransk{\'a} Lomnica, Slovak Republic}
\affil{Leibniz Institute for Astrophysics Potsdam (AIP),
    An der Sternwarte 16,
    14482 Potsdam, Germany}
\affil{University of Potsdam, Institute of Physics and Astronomy,
    Karl-Liebknecht-Str.\ 24/25,
    14476 Potsdam, Germany}
    
\author[0000-0002-9858-0490]{Andrea Diercke}
\affil{Leibniz Institute for Astrophysics Potsdam (AIP),
    An der Sternwarte 16,
    14482 Potsdam, Germany}
\affil{University of Potsdam, Institute of Physics and Astronomy,
    Karl-Liebknecht-Str.\ 24/25,
    14476 Potsdam, Germany}

\author[0000-0002-2366-8316]{Harry Enke}
\affil{Leibniz Institute for Astrophysics Potsdam (AIP),
    An der Sternwarte 16,
    14482 Potsdam, Germany}
    
\author[0000-0002-5883-4273]{Jochen Klar}
\affil{Leibniz Institute for Astrophysics Potsdam (AIP),
    An der Sternwarte 16,
    14482 Potsdam, Germany}

\author[0000-0002-4739-1710]{Horst Balthasar}
\affil{Leibniz Institute for Astrophysics Potsdam (AIP),
    An der Sternwarte 16,
    14482 Potsdam, Germany}
    
\author[0000-0001-5963-8293]{Rohan E.\ Louis}
\affil{Center of Excellence in Space Sciences India (CESSI), 
    Indian Institute of Science Education and Research Kolkata, 
    Nadia 741246, West Bengal, India}

\affil{Leibniz Institute for Astrophysics Potsdam (AIP),
    An der Sternwarte 16,
    14482 Potsdam, Germany}
        
\author[0000-0002-4645-4492]{Ekaterina Dineva}
\affil{Leibniz Institute for Astrophysics Potsdam (AIP),
    An der Sternwarte 16,
    14482 Potsdam, Germany}
\affil{University of Potsdam, Institute of Physics and Astronomy,
    Karl-Liebknecht-Str.\ 24/25,
    14476 Potsdam, Germany}

%
%

\begin{abstract}\noindent
In high-resolution solar physics, the volume and complexity of 
photometric, spectroscopic, and polarimetric ground-based data significantly
increased in the last decade reaching data acquisition rates of terabytes per
hour. This is driven by the desire to capture fast processes on the Sun and
by the necessity for short exposure times ``freezing'' the atmospheric seeing,
thus enabling post-facto image restoration. Consequently, large-format and
high-cadence detectors are nowadays used in solar observations to facilitate 
image restoration. Based on our experience during the ``early science'' phase 
with the 1.5-meter GREGOR solar telescope (2014--2015) and the subsequent
transition to routine observations in 2016, we describe data collection and 
data management tailored towards image restoration and imaging spectroscopy.
We outline our approaches regarding data processing, analysis, and archiving 
for two of GREGOR's post-focus instruments 
(see \href{http://gregor.aip.de}{gregor.aip.de}), i.e., the GREGOR Fabry-P\'erot
Interferometer (GFPI) and the newly installed High-Resolution Fast Imager
(HiFI). The heterogeneous and complex nature of multi-dimensional data 
arising from high-resolution solar observations provides an intriguing but 
also a challenging example for ``big data'' in astronomy. The big data 
challenge has two aspects: (1) establishing a workflow for publishing the 
data for the whole community and beyond and (2) creating a Collaborative 
Research Environment (CRE), where computationally intense data and 
post-processing tools are co-located and collaborative work is enabled for
scientists of multiple institutes. This requires either collaboration with 
a data center or frameworks and databases capable of dealing with huge data 
sets based on Virtual Observatory (VO) and other community standards and
procedures.
\end{abstract}


%
%

\section{Introduction}

Challenges posed by ``Big Data'' certainly became a topic in solar physics with 
the launch of space missions such as the Solar and Heliospheric Observatory 
\citep[SoHO,][]{Domingo1995} and the Solar Dynamics Observatory 
\citep[SDO,][]{Pesnell2012}. Synoptic full-disk images (visible, UV, and EUV), 
magnetograms, and Doppler maps are the core data products of both missions. 
However, SDO pushed the limits of spatial resolution to one second of arc and 
a cadence of 12~seconds. The results are about 300 million images and a 
total data volume of more than 3.5~petabytes during the seven year mission time 
so far. Dataflow and data processing for SDO are described in 
\citet{Martens2012} concerning both hard- and software aspects, and in 
particular the data conditioning for helioseismology and near real-time data
products needed in space weather prediction and forecast. Many of these 
data products are available from the Joint Science Operations 
Center\footnote{\href{http://jsoc.stanford.edu}{jsoc.stanford.edu}} (JSOC) at 
Stanford University. Considering the volume of data, various institutes around 
the world hold partial or full copies of SoHO and SDO for advanced in-house 
processing.

The data volume of ground-based synoptic full-disk observations also 
significantly increased over the years, even though not reaching the 
magnitude of space data. Both helioseismology and space weather applications
require continuous observations, thus telescope networks such as the  Global
Oscillation Network Group \citep[GONG,][]{Leibacher1999} and the  Global 
H$\alpha$ Network \citep{Denker1999a, Steinegger2000b} are a natural choice
to overcome the day-night cycle. Other important synoptic data sets are 
hosted at the ``Digital
Library''\footnote{\href{http://diglib.nso.edu}{diglib.nso.edu}} of the U.S.\
National Solar Observatory (NSO), e.g., the photospheric and chromospheric 
vector magnetograms of the Synoptic Optical Long-term Investigations of the
Sun \citep[SOLIS,][]{Keller2003, Henney2009} program, or at institutional 
data repositories, e.g., full-disk images of the Chromosheric
Telescope\footnote{\href{http://www.leibniz-kis.de/en/observatories/chrotel/data}{www.leibniz-kis.de/en/observatories/chrotel/data}}
\citep[ChroTel,][]{Kentischer2008} operated by the Kiepenheuer Institute 
for Solar Physics in Freiburg, Germany and coronal
images\footnote{\href{http://www2.hao.ucar.edu/mlso/mlso-data-and-movies}{www2.hao.ucar.edu/mlso/mlso-data-and-movies}} of the Coronal Multichannel Polarimeter
\citep[CoMP,][]{Tomczyk2008} and of the COSMO K-coronagraph \citep{Tomczyk2016}
from the Mauna Loa Solar Observatory. These data are typically accessible via
FTP archives or can be requested via web-based query forms.

High-spectral-resolution spectroscopy and spectropolarimetry were
historically the domain of ground-based telescopes. The Japanese Hinode
\citep{Kosugi2007} space mission with its 50-centimeter Solar Optical 
Telescope \citep[SOT,][]{Tsuneta2008} changed this field by providing
high-spectral and high-spatial resolution full Stokes polarimetry with
moderate temporal resolution but high sensitivity \citep{Ichimoto2008}. 
Hinode data are publicly available in the Data ARchive and Transmission 
System \citep[DARTS,][]{Miura2000} at Institute of Space and Astronautical 
Science (ISAS) in Japan and are also mirrored to data centers around the 
world \citep{Matsuzaki2007}.

Access to high-resolution ground-based data is often difficult because they 
were obtained in campaigns led by a principle investigator (PI) and his/her 
team. In addition, poor documentation and offline storage of 
high-resolution data (tape drives and local hard-disk drives) have hampered 
the efforts to make them openly accessible. Fortunately, the situation is
improving as more and more data holdings adopt the open-access paradigm. 
For example, high-resolution data from the 1.6-meter Goode Solar Telescope
\citep{Cao2010} at Big Bear Solar Observatory are listed on the
observatory website and can be requested via web-interface. In preparation 
for the next generation of large-aperture solar telescopes, NSO tested a 
variety of observing modes including proposal-based queue observations 
executed by experienced scientists and observatory staff. These service-mode
data\footnote{\href{http://nsosp.nso.edu/node/250}{nsosp.nso.edu/node/250}}
became also publicly available and are accessible via the NSO Digital 
Library infrastructure.

At the moment, the Daniel K.\ Inouye Solar Telescope
\citep[DKIST,][]{McMullin2014, Tritschler2016} is under construction with 
first light anticipated in 2020, and the European Solar Telescope
\citep[EST,][]{Collados2010a, Collados2010b} completed its preliminary
design phase and was included in the 2016 roadmap of the European Strategy 
Forum for Research Infrastructures (ESFRI). Concepts for the exploitation
of DKIST, with special emphasis on the ``Big Data'' challenge were presented 
in \citet{Berukoff2016}. The current generation of 1-meter-class solar
telescopes like GREGOR \citep{Schmidt2012} and the Goode Solar Telescope can be considered 
as stepping stones to unveil the fundamental spatial scales of the solar
atmosphere, i.e., the pressure scale height, the photon mean free path, and
the elementary magnetic structure size. All scales are of the order of 100
kilometers or even smaller. This provides the impetus for large-aperture 
solar telescopes and high-resolution solar physics, i.e., to approach 
temporal and spatial scales that are otherwise only accessible 
with numerical radiative MHD simulations \citep[e.g.,][]{Voegler2005, 
Rempel2014, Beeck2015}.

In this article, we introduce high-resolution ground-based imaging 
(spectroscopic) data obtained with the 1.5-meter GREGOR solar telescope 
and describe our approaches to data processing, analysis, management, 
and archiving. We show by example how the dichotomy of ground-based vs.\
space-mission data and synoptic vs.\ high-resolution data affects these
approaches. In addition, we present the collaborative research environment 
(CRE) for GREGOR data as a concept tailored towards the needs of the
high-resolution solar physics community. This is complementary to research
infrastructures such as the Virtual Solar
Observatory\footnote{\href{http://umbra.nascom.nasa.gov/vso}{umbra.nascom.nasa.gov/vso}, \href{http://vso.nso.edu/}{vso.nso.edu}, and
\href{http://vso.stanford.edu}{vso.stanford.edu}} \citep[VSO,][]{Hill2004c}, 
which was established to allow easy access of solar data from various space
missions as well as from ground-based observatories. The VSO reduces the 
effort of locating and downloading different data in different archives 
by providing a common web-based interface. The VSO design minimizes the hard-
and software resources of federated archives and often simplifies
integration of new data. The VSO does not store any data but integrates
federated archives by offering a central registry and interfaces for 
distributed queries to multiple independent data repositories. An alternative
access to solar data is provided by the SolarSoft \citep{Bentley1998}
library, which is written mainly in the Interactive Data
Language\footnote{\href{http://harrisgeospatial.com}{harrisgeospatial.com}} 
(IDL). Virtual observatory implementations in solar physics also exist on 
European level with the European Grid of Solar Observations 
\citep{Bentley2002} and more recently with the SOLARNET Virtual
Observatory\footnote{\href{http://solarnet.oma.be/}{solarnet.oma.be}} 
(SVO). Currently, only an SVO prototype is available. However, data can be
searched based on events, data set specific parameters, and co-temporal
observations.

In the following, we provide a comprehensive overview of GREGOR 
high-resolution data -- from the photons arriving at the detector to the 
final data products. In Sect.~\ref{SEC2}, we describe the telescope, its
post-focus instruments, and the particulars of high-cadence imaging. The 
data processing pipeline, its design, and its relation to other software 
libraries is introduced in Sect.~\ref{SEC3}. Data management and the access 
to the GREGOR GFPI and HiFI
archive\footnote{\href{http://gregor.aip.de}{gregor.aip.de}} (Sect.~\ref{SEC4}) 
comprise the domain specific answers to the challenges provided by 
``Big Data'' in solar and stellar astronomy. Finally, the conclusions in
Sect.~\ref{SEC5} develop a perspective for CREs in high-resolution solar 
physics and explore future extensions allowing database research.

%
%

\section{GREGOR Solar Telescope and Instrumentation}\label{SEC2}


\subsection{GREGOR Solar Telescope}\label{SEC21}

The 1.5-meter GREGOR solar telescope is the largest telescope in Europe 
for high-resolution solar observations (see \citet{Soltau2012} for the 
origin of the telescope's name and why it is formatted in capital letters).
Located at Observatorio del Teide, Iza\~na, Tenerife, Spain, the telescope
exploits the excellent and stable seeing conditions of a mountain-island
observatory site. The concept for the telescope's mechanical structure
\citep{Volkmer2012} and the open design employing a foldable-tent dome
\citep{Hammerschlag2012} allow for wind flushing of the telescope platform, 
which minimizes dome and telescope seeing. The GREGOR telescope uses a double
Gregory configuration \citep{Soltau2012} to limit the field-of-view (FOV) to 
a diameter of 150\arcsec\ to facilitate on-axis polarimetric calibrations in
the symmetric light path, and to provide a suitable $f$-ratio to the GREGOR
Adaptive Optics System \citep[GAOS,][]{Berkefeld2012} and the four post-focus
instruments: Broad-Band Imager \citep[BBI,][]{vonderLuehe2012}, GREGOR 
Infrared Spectrograph \citep[GRIS,][]{Collados2012}, GREGOR Fabry-P\'erot
Interferometer \citep[GFPI,][and references therein]{Denker2010a, 
Puschmann2012}, and High-Resolution Fast Imager \citep[HiFI,][]{Denker2018c}. 
In the present study, we discuss data processing, analysis, management, and
archiving for the last two instruments.


\subsection{GREGOR Fabry-P\'erot Interferometer}\label{SEC22}

The GFPI is a tunable, dual-etalon imaging spectrometer, where the etalons are 
placed near a conjugated pupil plane in the collimated beam. The etalons were 
manufactured by IC Optical Systems (ICOS) in Beckenham, UK. They have a 
70-millimeter diameter free aperture and possess both high finesse (${\cal 
F}_\mathrm{eff} = 45$\,--\,50) and high reflectivity ($R \approx 95$\%). Their 
coatings are optimized for the spectral range 5300\,--\,8600~\AA, where the 
instrument achieves a spectral resolution of ${\cal R} \approx 250\,000$. 
Spectral scans are recorded with two precisely synchronized Imager QE CCD 
cameras from LaVision in G\"ottingen, Germany. One camera acquires narrow-band 
filtergrams, whereas the simultaneous broad-band images enable image restoration 
of the full spectral scan using various deconvolution techniques.

The 12-bit images with $1376 \times 1040$ pixels are captured at a rate of up 
to 10~Hz for full frames depending on exposure time. The image scale of about 
0.04\arcsec\ pixel$^{-1}$ yields a FOV of $55\arcsec \times 42\arcsec$. The 
image acquisition rate can be doubled with 2$\times$2-pixel binning. The 
relatively small full-well capacity of 18\,000~e$^-$ and low readout noise of 
the detectors are well adapted to the instrument design considering the small 
full-width-at-half-maximum (FWHM) of the double-etalon spectrometer of 
just 25\,--\,40~m\AA, the maximum quantum efficiency of 60\% at 5500~\AA,
and the short exposure times ($t_\mathrm{exp} = 10$\,--\,30~ms) needed to
``freeze'' the seeing-induced wavefront distortions. The typical cadence 
of $\Delta t = 20$\,--\,60~s for a spectral scan provides very good 
temporal resolution so that dynamic processes in the solar photosphere 
and chromosphere can be resolved. 

However, both exposure time and cadence are already compromises because 
the high-spectral resolution leads to a low number of incident photons at the
detector, and small-scale features potentially move or evolve significantly in
the given time interval. Faster detectors combining low readout noise,
comparatively small full-well capacity, good quantum efficiency, and a high 
duty cycle with respect to the exposure time mitigate against these limitations
and are considered for future upgrades. In principle, the GFPI can be operated
in a polarimetric mode, which produces spectral scans of the four Stokes
parameters so that the magnetic field vector can be inferred for each pixel.
However, validation of the polarimetric mode is still in progress. 
Consequently, imaging polarimetry is not covered in this study.

\renewcommand{\baselinestretch}{0.85}\normalsize
\begin{figure*}[t]
\includegraphics[width=\textwidth]{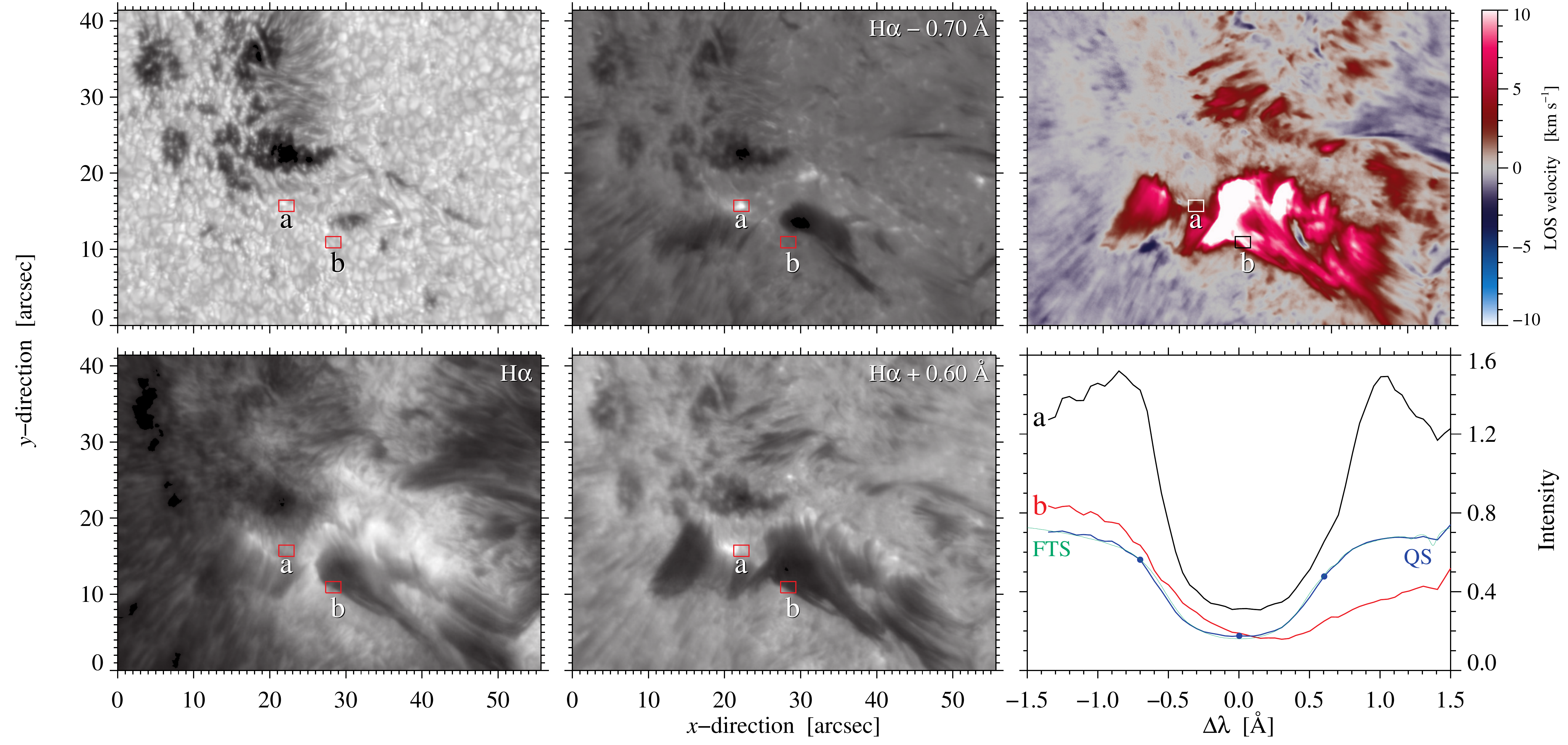}
\caption{Imaging spectroscopy of active region NOAA 12139 observed
    on 2014 August~14 with the GFPI in the strong chromospheric absorption
    line H$\alpha$ $\lambda$6562.8~\AA: broad-band image (\textit{top-left})
    near the H$\alpha$ spectral region, line-core intensity image
    (\textit{bottom-left}), blue line-wing image at $\Delta \lambda =
    -0.70$~\AA\ (\textit{top-middle}), red line-wing image at $\Delta \lambda
    =+0.60$~\AA\ (\textit{bottom-middle}), and chromospheric Doppler velocity
    derived with the Fourier phase method (\textit{top-right}), where blue 
    and red colors represent up- and downflows, respectively. Samples of 
    spectral profiles (\textit{bottom-right}) are shown for areas `a' 
    (\textit{black}), `b' (\textit{red}), quiet-Sun `QS' (\textit{blue}), 
    and Fourier Transform Spectrometer \citep[FTS,][]{Wallace1998} spectral 
    atlas (\textit{green}). The bullets on the quiet-Sun profile denote the positions of the three displayed filtergrams.}
\label{FIG01}
\end{figure*}
\renewcommand{\baselinestretch}{1.0}\normalsize

Sample data products obtained from a scan of the strong chromospheric
absorption line H$\alpha$ $\lambda$6562.8~\AA\ are compiled in Fig.~\ref{FIG01} 
to illustrate the GFPI's science capabilities. Ellerman bombs \citep[see][for
a review]{Rutten2013}, as opposed to micro-flares, are typically observed as
enhanced line-wing emission in H$\alpha$, H$\beta$, H$\gamma$, etc.\ and have
typical lifetime of 1.5\,--\,7~min with a maximum of around 30~min
\citep{Pariat2007}. To resolve their evolution requires fast spectral scans 
with imaging spectroscopy and post-facto image restoration to uncover 
small-scale dynamics within the photosphere and chromosphere. On 2014 
August~14, active region NOAA~12139 was observed with the GFPI focusing on a
complex group of sunspots and pores (see broad-band image in Fig.~\ref{FIG01}).
Image restoration using Multi-Object Multi-Frame Blind Deconvolution 
\citep[MOMFBD,][]{Loefdahl2002, vanNoort2005}
was applied to the spectral data to enhance solar fine-structure.
The blue and red line-wing filtergrams reveal two features with different spectral
characteristics: an ``Ellerman bomb'' \citep{Ellerman1917} as localized, 
small-scale brightenings (area `a') and a system of dark fibrils with strong 
downflows associated with newly emerging flux (area `b'), respectively. 

High-spatial resolution imaging and imaging spectroscopy belong to
the standard observational techniques of large-aperture solar telescopes. 
Various studies based on GFPI data illustrate the potential of the instrument.
Recently, \citet{Kuckein2017b} used GFPI Ca\,\textsc{ii} 8542.1~\AA\
filtergrams to study sudden chromospheric small-scale brightenings. The
combination of ground-based, high-resolution imaging spectroscopy and 
synoptic EUV full-disk images from space reveals that the brightenings 
belong to the footpoints of a micro-flare. To further investigate the 
bright kernels and the central absorption part (below the flaring arches),
spectral inversions of the near-infrared Ca\,\textsc{ii} line  are performed 
with the NICOLE code \citep{SocasNavarro2015}. The retrieved average 
temperatures reveal rapid heating at the brightenings (footpoints) of 
the micro-flare of about 600~K. The inferred line-of-sight (LOS) velocities
at the central absorption area show upflows of about $-2$~km~s$^{-1}$. In 
contrast, downflows dominate at the other footpoints.

\begin{figure}[t]
\includegraphics[width=\columnwidth]{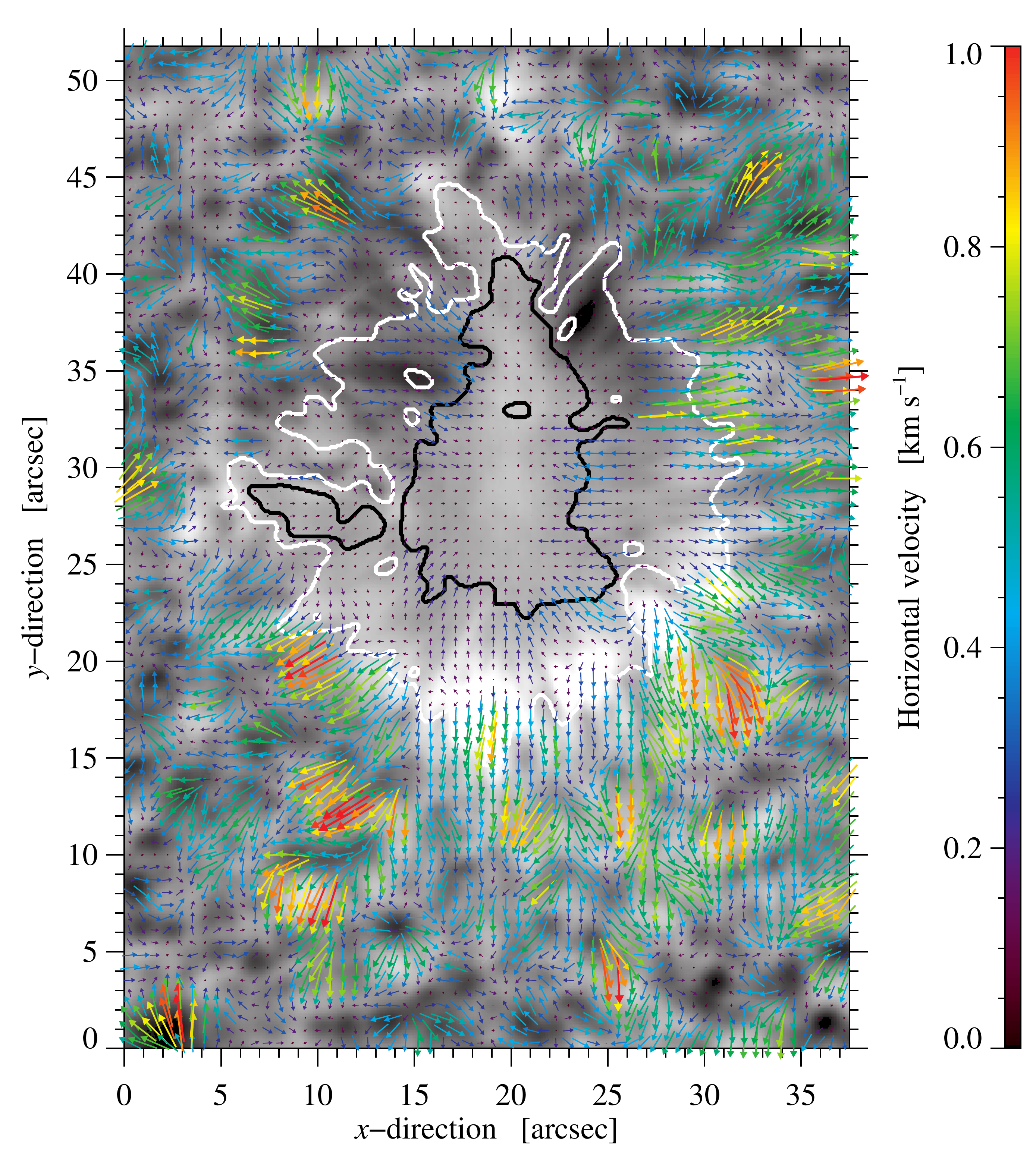}
\caption{Three-dimensional flow field observed in active region NOAA~12597 
     on 2016 September~24. The horizontal flows were derived from a time-series 
     of restored GFPI broad-band images at $\lambda$6122.7~\AA, whereas the
     Doppler velocities were determined from a restored narrow-band scan of 
     the photospheric Fe\,\textsc{i} $\lambda$6173.3~\AA\ line. Color-coded 
     local correlation tracking \citep[LCT,][]{November1989, Verma2011} vectors are superposed onto the average LOS velocity map, which was
     scaled between $\pm 0.7$~km~s$^{-1}$. The black and white contours 
     delineate the umbra-penumbra and penumbra-granulation boundaries,
     respectively.}
\label{FIG02}
\end{figure}

In another study, \citet{Verma2018} used high-resolution imaging 
and spectroscopic GFPI data in the photospheric Fe\,\textsc{i}
$\lambda$6173.3~\AA\ spectral line to infer the three-dimensional velocity
field associated with a decaying sunspot penumbra (Fig.~\ref{FIG02}). 
The velocities in the decaying penumbral region deviate from the usual 
penumbral flow pattern because of flux emergence in the vicinity of the 
sunspot. The detailed analysis is based not only on GFPI data, but includes 
HiFI and GRIS observations, which provide further photospheric and 
chromospheric diagnostics. In a study, with a similar set-up for GREGOR
multi-wavelength and multi-instrument observations, \citet{Felipe2017}
investigated the impact of flare-ejected plasma on sunspot fine-structure, 
i.e., a strong light-bridge, which experienced localized heating and changes 
of its magnetic field structure.


\subsection{High-Resolution Fast Imager}\label{SEC23}

Detectors based on scientific Complementary Metal-Oxide-Semiconductor (sCMOS) 
technology have become an alternative to standard CCD devices in
astronomical applications, in particular when high-cadence and large-format
image sequences are needed. In early 2016, we installed HiFI at the GREGOR 
solar telescope, where it observes the blue part of the spectrum 
(3850\,--\,5300~\AA) using a dichroic beamsplitter in the GFPI's optical path. 
Two Imager sCMOS cameras from LaVision \citep{LaVision2015b} are synchronized by a 
programmable timing unit and record time-series suitable for image restoration 
either separately for each channel or making use of both channels at the same time. 

\renewcommand{\baselinestretch}{0.85}\normalsize
\renewcommand{\baselinestretch}{1.0}\normalsize

\begin{figure}[t]
\includegraphics[width=\columnwidth]{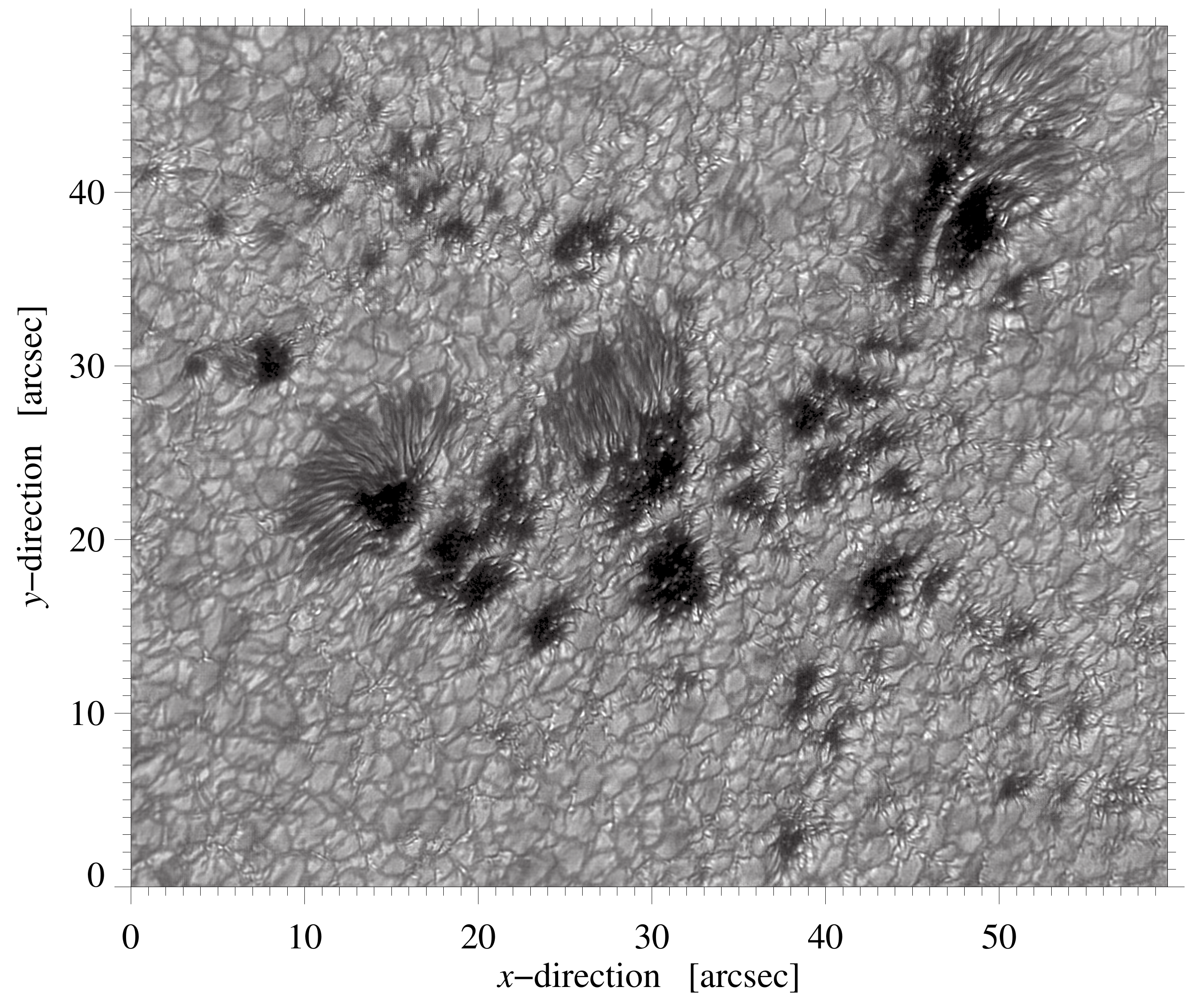}
\caption{Blue continuum image $\lambda$4505.5~\AA\ of active region NOAA 12529 
     obtained with HiFI at 08:37~UT on 2016 April~11. The image was restored
     from a time-series of 100 short-exposure images with the speckle masking
     method implemented in KISIP.}
\label{FIG03}
\end{figure}

In typical observations, two of three spectral regions are selected, i.e., 
Ca\,\textsc{ii}\,H $\lambda$3968.0~\AA, Fraunhofer G-band $\lambda$4307.0~\AA, 
and blue continuum $\lambda$4505.5~\AA\ (see Fig.~\ref{FIG03}). The width 
of the filters is around 10~\AA, so that typical exposure times reach from a
fraction of a millisecond to a few milliseconds. Thus, observations are not
``photon-starved'' as often encountered in imaging spectropolarimetry, where 
the transmission profile of (multiple) Fabry-P\'erot etalons can be as narrow 
as 25~m\AA. Different count rates in both channels are balanced by choosing
suitable neutral density filters. Short exposure times are essential to freeze 
the wavefront aberrations in a single exposure. The 2560$\times$2160-pixel 
images with a FOV of $64.8\arcsec \times 54.6\arcsec$ are 
digitized as 16-bit integers and recorded with a data acquisition rate of 
almost 50~Hz. Thus, the image scale is about 0.025\arcsec~pixel$^{-1}$ or about 
18~km on the solar surface at disk center. The diffraction-limited 
resolution of the GREGOR telescope with a diameter $D = 1.5$~m at a wavelength 
$\lambda = 4000$~\AA\ is $\alpha = \lambda / D = 0.055\arcsec$. According to the 
Nyquist sampling theorem, HiFI images are critically sampled at the shortest 
wavelength of the standard interference filters.

\renewcommand{\baselinestretch}{0.85}\normalsize
\renewcommand{\baselinestretch}{1.0}\normalsize

In the standard HiFI observing mode, sets of 500 images are captured in 
each channel in 10~s and continuously written to a RAID-0 array of SSDs at 
a cadence below 20~s. The imaging system has achieved a sustained write 
speed summed over both channels of up to 660~MB~s$^{-1}$. To limit the final 
data storage requirements, only
the best $2 \times 100$ images of a set are kept for image restoration and 
further data analysis. These settings are already a compromise considering 
that solar features move with velocities of several kilometers per second in 
the photosphere and several tens of kilometers per second in the chromosphere,
often exceeding the speed of sound. Eruptive phenomena in the chromosphere 
reach even higher velocities in excess of 100~km~s$^{-1}$. Thus, in the 
standard observing mode, solar features moving at more than 2~km~s$^{-1}$ 
and traversing a pixel in less than about 10~s will be blurred, and their 
proper motions will not be properly resolved. However, considering that 
HiFI provides mainly context data and that only some pixels are exposed to 
high velocities, this compromise is acceptable. If higher temporal resolution
is required, for example when tracking small-scale bright points or following 
the evolution of explosive events, changing the observing mode is an option, 
i.e., reading out smaller sub-fields to increase the data acquisition rate or
dropping frame selection to keep all observed frames. 

\renewcommand{\baselinestretch}{0.85}\normalsize
\begin{figure*}[t]
\includegraphics[width=\textwidth]{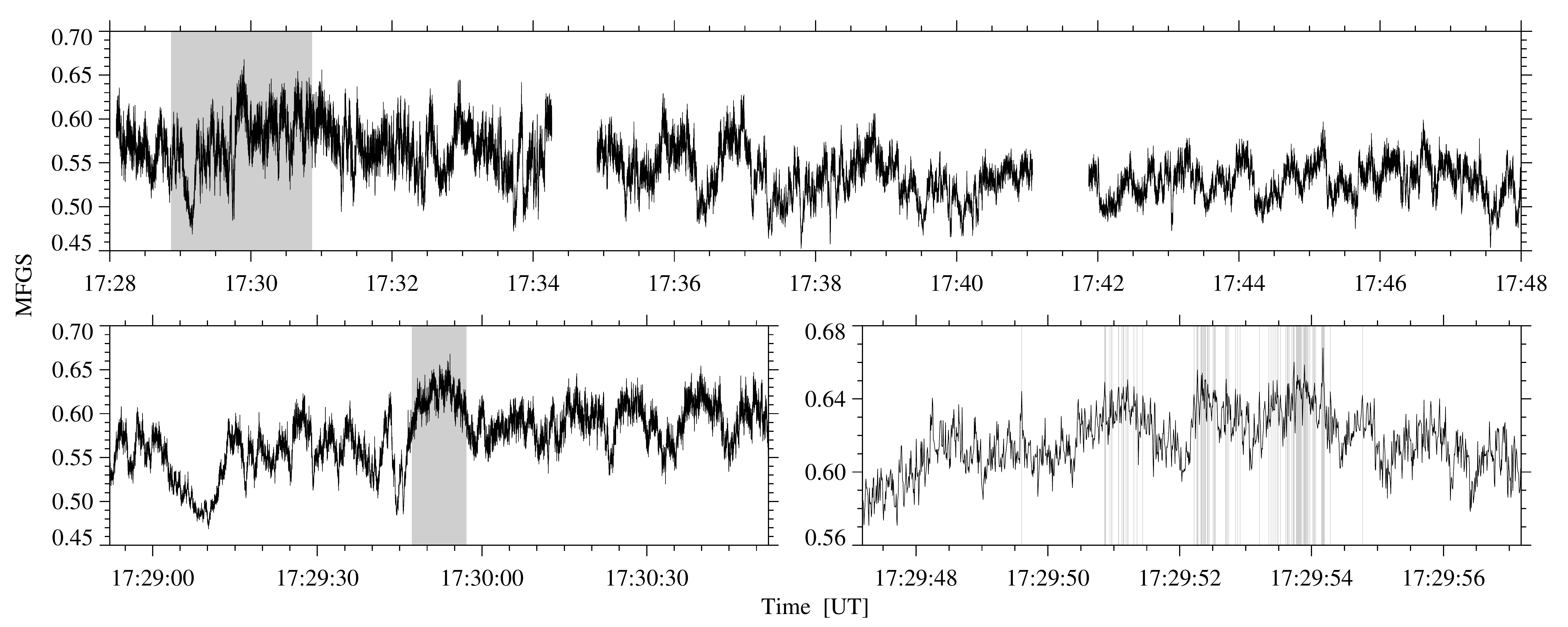}
\caption{Temporal evolution of the seeing conditions at the GREGOR solar 
    telescope on 2017 March~25. Three series of 50\,000 images were captured at 
    135~Hz (\textit{top}), and the corresponding MFGS values were computed and
    are plotted for G-band images. The gray rectangle covers a 2-minute period,
    which is depicted at higher resolution (\textit{bottom-left}). A 10-second
    period at even higher temporal resolution highlights the frame selection
    process (\textit{bottom-right}). The best images, which are used for 
    image restoration, are marked by gray vertical lines.}
\label{FIG04}
\end{figure*}
\renewcommand{\baselinestretch}{1.0}\normalsize


\subsection{Implications of High-Cadence Imaging}\label{SEC24}

As demonstrated above, many instrument designs in solar physics rely on 
high-cadence imaging. The data challenge arises from the combination of 
several factors: the daytime correlation time-scale of the seeing, the 
evolution time-scale of solar features, and large-format detectors. 
The latter are needed to either catch transient events or to observe
large-scale, coherent features that provide structuring and connectivity 
in the solar atmosphere. To illustrate the implications, we carried out 
a small experiment with the HiFI cameras. Images were acquired at 135~Hz 
for only a small FOV of $640 \times 640$ pixels, i.e., $16.3\arcsec \times
16.3\arcsec$ on the solar disk. Three sets of 50\,000 images were written
to disk with short interruptions at a sustained rate of 220~MB~s$^{-1}$. 
An extended study based on an even higher image acquisition rate 
is presented in \citet{Denker2018b}, where we evaluated image quality metrics
and the impact of frame selection for AO-corrected images on image restoration
with the speckle masking technique \citep{Weigelt1983, vonderLuehe1993, deBoer1993}.

\renewcommand{\baselinestretch}{0.85}\normalsize
\begin{figure}[t]
\includegraphics[width=\columnwidth]{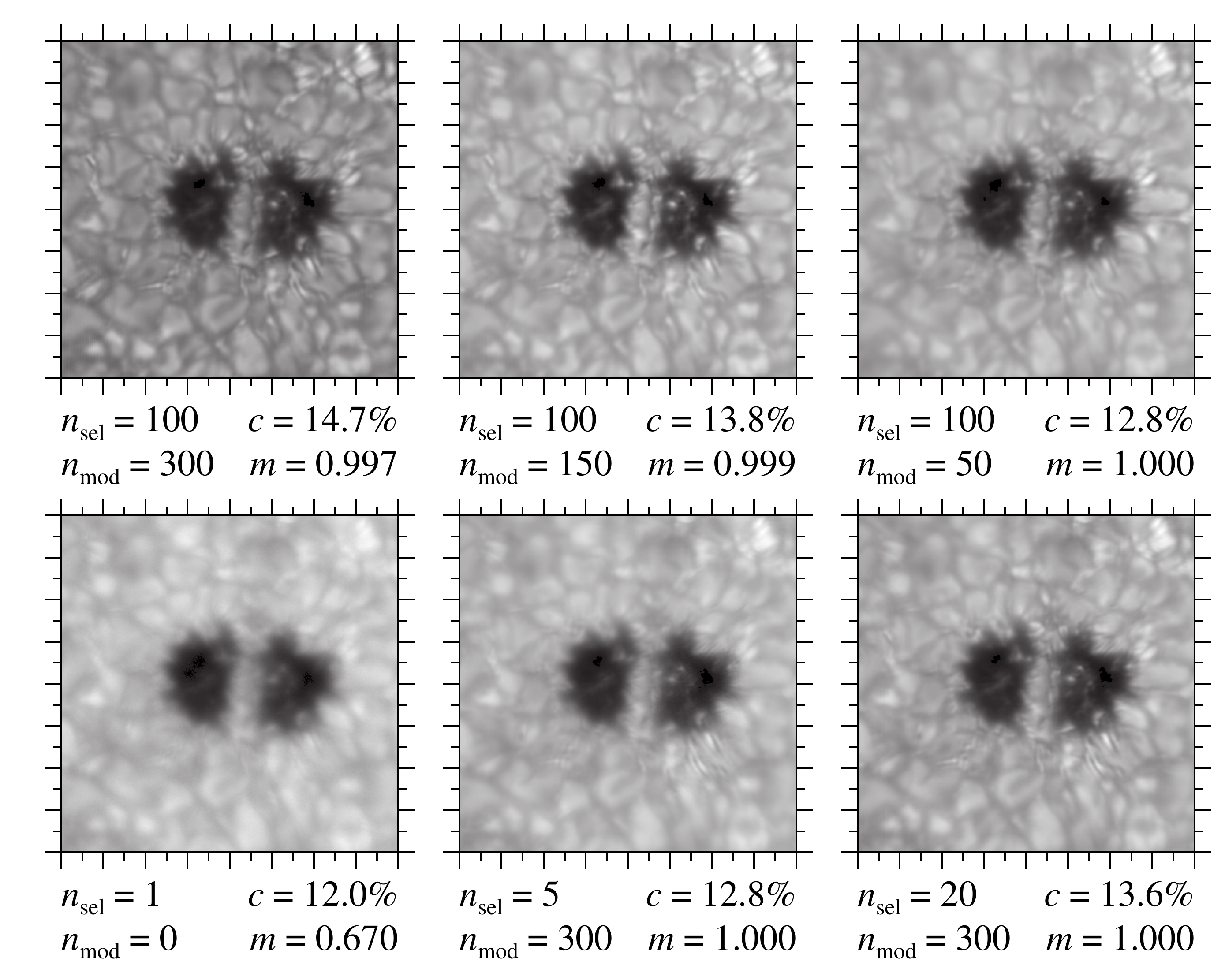}
\caption{Decaying pore with light-bridge observed in active region NOAA~12643 on 
    2017 March~25. All G-band images were restored with MFBD with the exception 
    of the best raw image (\textit{bottom-left}). The best $n_\mathrm{sel}$ 
    images from a 10-second time interval were used for the restored image, 
    which is based on $n_\mathrm{mod}$ restored Zernike modes. The 
    rms-intensity contrast $c$ and the MFGS value $m$ are used as image 
    quality metrics. The images are scaled individually between minimum and
    maximum intensity, and the FOV is $16\arcsec \times 16\arcsec$.}
\label{FIG05}
\end{figure}
\renewcommand{\baselinestretch}{1.0}\normalsize

The median filter gradient similarity \citep[MFGS,][]{Deng2015} is an 
image quality metric recently introduced into solar physics. The results 
for G-band images of the full time-series are depicted in the top panel 
of Fig.~\ref{FIG04}, whereas the two lower panels show successively shorter
time periods centered around the moment of best seeing conditions. The
AO system locked on a decaying pore in active region NOAA~16643, which 
contained a light-bridge, umbral dots, and indications of penumbra-like
small-scale features, i.e., elongated, alternating bright and dark features
at the umbra-granulation boundary as well as chains of filigree in the
neighboring quiet Sun, which point radially away from the pore's center
(Fig.~\ref{FIG05}). In the standard HiFI observing mode, $2 \times 100$ 
images are selected within a 10-second period as indicated by the gray 
vertical bars in the lower-right panel of Fig.~\ref{FIG04}. This 
observing mode relies on a special implementation of frame selection
\citep{Scharmer1989, Scharmer1991, Kitai1997}, where not just the best 
solar image is selected but a set of high-quality images is chosen for 
image restoration.

Obviously, even at this very high cadence, the image quality metrics show 
still strong variations, despite some clustering of the best images in
Fig.~\ref{FIG04}, indicating that seeing fluctuations occur at even higher
frequencies. Considering what lies ahead for high-resolution imaging, 
the data acquisition rates for the next generation of $4\mathrm{k} \times 
4\mathrm{k}$-pixel detectors recording at 100~Hz will amount to about 
3~GB~s$^{-1}$. This exceeds today's typical data acquisition rates of less 
than 1~GB~s$^{-1}$, but demonstrates that the data challenge persists for 
the years to come and in particular for the next generation of large-aperture 
solar telescopes such as DKIST and EST. In \citet{Denker2018b} we demonstrated 
that an image acquisition rate of $f_\mathrm{acq} = 50$~Hz is an appropriate
choice, considering the marginal benefits in quality for the frame-selected 
images with respect to the demands on camera detectors, network bandwidth, 
data storage, and computing power.

Multi-frame blind deconvolution \citep[MFBD,][]{Loefdahl2002} is 
another commonly used image restoration technique in solar physics.
Figure~\ref{FIG05} compiles some results based on the best set of G-band 
images. In \citet{Denker2018b}, we established a benchmark for G-band 
images, i.e., an MFGS value of $m = 0.65$, where image restoration with the 
speckle masking technique becomes possible. Thus, the seeing conditions were 
only moderate to good on 2017 March~25. The aforementioned threshold was
determined for the full FOV of the sCMOS detector, whereas the present 
observations used a much smaller FOV covering the immediate neighborhood of the
AO lock point. The top row of Fig.~\ref{FIG05} demonstrates that even
compared to telescopes with 0.7\,--\,1 meter apertures, data obtained with 
larger 1.5-meter-class telescopes require a significant increase in the number 
of Zernike modes ($n_\mathrm{mod} \approx 300$) in the restoration process.
This significantly increases the computation time and poses challenges for 
imaging with 4-meter-class solar telescopes. MFBD has an advantage over 
speckle masking because it requires a smaller number of images for a 
restoration, in particular, when the images are of high quality taken 
under very good or excellent seeing conditions. Thus, restored time-series
with higher cadence become possible. Already, $n_\mathrm{sel} = 20$ selected
images deliver good restorations. Note that MFGS looses its discriminatory
power for restored images so that other metrics like the rms-intensity 
contrast $c$ become more important. A likely explanation is that the 
MFGS metric is sensitive to the fine-structure contents of an image, which is 
mainly encoded in the phases of the Fourier-transformed image. Thus, once almost
diffraction-limited information is recovered, the MFGS metric reaches a plateau. 
The image contrast on the other hand is more closely related to the Fourier
amplitudes.

%
%

\section{sTools Data Pipeline}\label{SEC3}

The ``Optical Solar Physics'' research group at AIP operates with GFPI and HiFI 
two of GREGOR's facility instruments. The software package ``sTools'' \citep[see 
][for a brief introduction]{Kuckein2017a} provides, among other features, the 
data processing pipeline for these two instruments. Major parts of the software 
were developed from 2013\,--\,2017 within the 
SOLARNET\footnote{\href{http://www.solarnet-east.eu}{www.solarnet-east.eu}} 
project, which is a ``Research Infrastructures for High-Resolution 
Solar Physics'' program following the Integrated Infrastructure Initiative 
(I3) model supported by the European Commission's FP7 Capacities Program. The 
software package was written from scratch but builds on the code development for 
and experiences gained with data reduction tools for the G\"ottingen 
Fabry-P\'erot Interferometer \citep{Bendlin1992, Puschmann2006b, Bello2008} and 
the Interferometric BIdimensional  Spectropolarimeter 
\citep[IBIS,][]{Cavallini2006}.

sTools is mainly written in IDL and utilizes other IDL libraries with robust and 
already validated programs whenever available. The SolarSoftWare (SSW) system 
\citep{Bentley1998, Freeland1998}, for example, offers instrument specific 
software libraries and utilities for ground-based instruments and space 
missions, which includes database access and powerful string processing for 
metadata. The MPFIT \citep{Markwardt2009} package is primarily used for spectral 
line fitting, the Coyote Library for image processing, graphics, and data I/O 
\citep{Fanning2011}, and the NASA IDL Astronomy User's 
Library\footnote{\href{http://idlastro.gsfc.nasa.gov}{idlastro.gsfc.nasa.gov}} 
for reading and writing data in the Flexible Image Transport System 
\citep[FITS,][]{Wells1981, Hanisch2001} format. Data from imaging 
spectropolarimetry benefits especially from the possibility of writing FITS 
image extensions with individual headers \citep{Ponz1994}, i.e., polarization 
state, wavelength position, and calibration information are saved along with the 
corresponding filtergram. The result is a compact (a few gigabytes), 
self-describing data set, which can serve as input for various spectral 
analysis and inversion codes. 

Computationally intense applications, in particular image restoration, make use 
of parallel computing implemented in other programming languages. Here, the IDL 
programs typically condition the input data and collect the output data for 
further processing. Currently, sTools provides interfaces for the 
Kiepenheuer-Institute Speckle Interferometry Program 
\citep[KISIP,][]{Woeger2008a} and MOMFBD. Recently, we started making use 
of the IDL built-in functions for parallel processing, so that other time 
consuming parts of the data processing are also more efficiently implemented.

All newly developed IDL routines use the prefix {\small\verb|stools_|} to avoid 
name space collisions with other libraries. No sub-folders exist within 
sTools with an exception for documentation and individual IDL routines from 
external sources, which are not part of the aforementioned libraries. 
Instrument-specific and functional dependencies are declared in the naming 
schema for routines, e.g., {\small\verb|stools_gfpi_|} for GFPI data processing 
or {\small\verb|stools_html_|} for creating summary webpages for the observed 
data sets. In general, we aim to separate instrument-specific code from 
multi-purpose routines, which can be shared among applications. The majority of 
the sTools programs are independent of the computer hardware and the site where 
the software is installed. Site specifications are encapsulated in structures, 
which are defined in specific configuration routines ({\small\verb|stools_cfg_|}). 
The same applies to expert knowledge, e.g., about spectral lines, filter 
properties, camera settings, telescope details, etc., which is collected 
in configuration routines that return structures with the required parameters 
based on tag names. For example, many fitting routine rely on information
regarding the width of spectral line and the spectral sampling. Based on the
configuration information the most suitable and validated fit parameters are
chosen. The configuration routines are the only programs, which may have to be 
adapted for specific sites, or if new observing modes are carried out with 
different filters and spectral lines. 

While writing the sTools routines, we placed special emphasis on proper 
documentation using standard IDL headers, which can be extracted with the 
{\small\verb|doc_library|} procedure, on meaningful inline comments, on 
descriptive variable names, and on a consistent programming style and 
formatting. Testing routines for specific data processing steps is
typically performed for several data sets with different observing
characteristics. This ensures that new calibration steps or updated procedures 
do not lead to unintended consequences. In some cases, when implementing new
observing set-ups for scanning multiple lines or using non-equidistant line
sampling, major conceptual changes of the code became necessary. These
test data are used to ensure that previously reduced and calibrated data is 
unaffected by the aforementioned changes. If changes affect already calibrated 
data, they will be reprocessed. The version of sTools is tracked so that, if 
needed, data calibrated with older versions can be recovered. These changes 
are documented on the project website and major updates will be accompanied 
by data release publications in scientific journals. Initial data 
calibration for one data set takes about one day computation time on a 
single processor. On the other hand, reprocessing all GFPI quick-look data, which
utilizes the already calibrated data, takes just a few days. The latter data serve 
as a benchmark to validate the performance of new or updated programs. The 
sTools data pipeline is currently installed at AIP and at the German solar
telescopes on Tenerife. The source code is maintained internally on an Apache 
Subversion\footnote{\href{http://subversion.apache.org}{subversion.apache.org}}
(SVN) version control system. The latest released version can be 
downloaded by registered users as a tarball from  the GREGOR webpages at AIP
(\href{http://gregor.aip.de}{gregor.aip.de}).

%
%

\section{Data Management and Data Archive}\label{SEC4}


\subsection{Point of Departure}\label{SEC41}

High-resolution solar observations are taken at 1-meter-class, 
ground-based solar telescopes around the world and with Hinode/SOT from 
space, where the latter provides in many respects (i.e., imaging and 
spectropolarimetry) the closest match to HiFI and GFPI data. 
Consequently, scientists, who work with such high-resolution data, are 
among the primary target groups for utilizing GFPI and HiFI data. In addition, 
we especially foresee close interactions with the solar physics community on 
spectral inversion codes and numerical modelling of the highly dynamic processes 
on the Sun. Therefore, our immediate priority is providing a CRE fostering
collaborations among researchers with common interests. Our goal is to
raise awareness of and stimulate interest in complex and heterogeneous
spectropolarimetric data sets, which are inherent trademarks of imaging
spectropolarimeters with a broad spectrum of user-defined observing sequences.
Offering a data repository to the whole solar physics community or the general
public at large will enhance the impact of these high-resolution data products.
However, in this case, it may be advantageous to highlight key data sets 
obtained under the best seeing conditions or of particularly interesting events
like solar flares. In any case, data access, as described in the data policy
(Sect.~\ref{SEC42}), is granted to everyone who registers for the GREGOR GFPI and 
HiFI data archive. At a later stage, when the high-level data products have matured,
a more differentiated access will be instantiated, dropping the registration
requirement for the publicly available data.  

The data specific challenges for GFPI and HiFI are:
\begin{itemize}
\item[$-$] The campaign- and PI-oriented nature of the data, which results 
    in different combinations of post-focus instruments with changing set-up 
    parameters, e.g., selection of diverse spectral lines, dissimilar spectral 
    and spatial sampling, and different cadences.
\item[$-$] Observations through Earth's turbulent atmosphere, which brings 
    about data with fast-changing quality and necessitates image restoration, 
    where different restoration algorithms introduce multiplicity in the major 
    data levels. 
\item[$-$] The complexity of data sets, which requires considerable efforts to
    condition the data products for a broader user base -- even at the level
    of quick-look data.
\item[$-$] The availability of personnel and financial resources for 
    maintaining long-term access to data and for quality assurance beyond the 
    typical funding cycle of third-party financial support. 
\end{itemize}    
Fortunately, AIP's mission includes the development of research technology and 
e-infrastructure as a strategic goal. Thus, the collaboration between E-Science 
and Solar Physics allowed us to develop a tightly matched solution to the 
aforementioned data specific challenges.


\subsection{Data Policy}\label{SEC42}

The GREGOR consortium (i.e., Kiepenheuer Institute for Solar Physics, 
Max Planck Institute for Solar System Research, and Leibniz Institute for 
Astrophysics Potsdam) and GREGOR partners (i.e., Instituto de Astrof\'{\i}sica 
de Canarias and Astronomical Institute of the Academy of Sciences 
of the Czech Republic) agreed that in principle all data shall be publicly 
available. In the following, we use the term ``consortium'' when referring to 
GREGOR members and partners.

Since observing proposals contain proprietary information 
and original ideas of the PI and her/his team, the data of PI-led observing 
campaigns will be embargoed for one year. The embargo period can be extended 
upon request for another year, when observations are related to PhD theses. 
Quick-look data are publicly available after storage at the GREGOR GFPI 
and HiFI archive and subsequent processing, typically after 4\,--\,6
weeks, and these data are not subject to the embargo period. This way, the PI 
of an observing can be contacted to inquire about potential collaborations even
within the embargo period. The consortium encourages, but does not require, 
collaboration with the originator of the data. All work based on GREGOR data is 
required to include an acknowledgement (see below), and the consortium asks the 
authors to include appropriate citations to the GREGOR reference articles 
published in 2012 as a special issue of \textit{Astronomische Nachrichten} 
(Vol.~333/9). A detailed version of the data policy is publicly available on 
the GREGOR webpages at AIP.


\subsection{GREGOR GFPI and HiFI Data Archive}\label{SEC43}

The archive is based on the Daiquiri\footnote{\href{http://escience.aip.de/daiquiri}{
escience.aip.de/daiquiri}} framework, which was developed by AIP's R\&D group
``Supercomputing and E-Science''. Daiquiri is designed for creating highly
customized web applications for data publication in astronomy. The features of
of Daiquiri comprise rich tools for user management, an SQL query interface to enable users to directly enter
database queries via the webpage, means to download the results of queries in
different formats following standards of the International Virtual Observatory
Alliance\footnote{\href{http://www.ivoa.net}{www.ivoa.net}}
\citep[IVOA,][]{Quinn2004}, and plotting functions. The interaction with
Daiquiri can be scripted using its VO Universal Worker Service
\citep[UWS,][]{Harrison2016} interface.

For the GREGOR archive, we use Daiquiri's user management to implement a custom
user-registration workflow. New users first register on the
GREGOR data portal, and they do not have to belong
to the GREGOR consortium. Therefore, registration is possible for anyone
with an interest in GREGOR high-resolution data. To emphasize the intended
collaborative nature of this research infrastructure, we use in the following
the terms ``GREGOR collaboration'' or ``collaborators'' collectively for all registered users. After the registration, one of the project managers confirms the new user,
again through the GREGOR portal. Only after this organizational confirmation
took place, members of the technical staff will activate the user. This
procedure is necessary because users receive CRE privileges, allowing them
processing data on institute-owned computers and editing of webpages and blog
entries.

Along with their accounts for the web portal, users also obtain
Secure Shell (SSH) login credentials for the data access node
(Sect.~\ref{SEC44}). The SSH protocol facilitates efficient and fast
distribution of the data products, while ensuring modern security and
maintainability (in particular firewall configuration), when comparing to the
popular, but insecure File Transfer Protocol (FTP). On the data access nodes, an elaborate permissions system ensures access restrictions and data security.
This is implemented using Linux Access Control Lists (ACLs), which extend
the usual file permissions (user/group/world) common on UNIX systems. Users are
organized in groups, which gain write permissions to certain sub-directories of
the GREGOR archive. These ACL directories, which typically comprise data for
a specific instrument, data processing level, and observing day, can have
multiple groups, allowing for fine-grained access by the registered users
(e.g., when accessing embargoed data) as well as the GREGOR archive administrators.

Besides the user-registration workflow, Daiquiri is also used to set up the GREGOR
webpages, some with access restrictions, some public. This includes the data products generated by sTools data processing
pipeline (see Sect.~\ref{SEC3}). Access to the data was initially limited to the GREGOR consortium but
is now open to all registered users of the GREGOR GFPI and HiFI data
archive. The data sets generated by sTools are currently in the process of being converted
to FITS files containing image extensions. This reduces the level of complexity, as
metadata are available and many images of a spectral scan or in a time-series are
already aggregated. Once converted to FITS format, these data will be integrated
into an SQL database and can be accessed using Daiquiri's query functionality.


\subsection{Collaborative Research Environment}\label{SEC44}

The large number and high-resolution of images and spectra, as well as the 
computational effort for their post-processing, demands capable and efficient 
structures for storage and data management. To make best use of GREGOR data
and to encourage their usage, we implemented
a dedicated CRE at AIP. This research infrastructure acted initially as 
central hub for storage and processing of different data products as well as 
their distribution within the GREGOR consortium but it is now open to all 
interested scientists. The CRE provides data space and data access with 
different levels of authorization, in addition to computational resources and 
customized tools for analysis and processing. Participation in the CRE is
managed by the GREGOR consortium lead for GFPI and HiFI. Finally,
collaborators have the option via the CRE to publish selected and curated
``science-ready'' data for the solar community, including a minted DOI registered
with DataCite.\footnote{\href{https://www.datacite.org}{www.datacite.org}}

Over the last decade, AIP provided similar CREs for several projects. For 
collaborations working on simulations of cosmological structure formation such 
as Constrained Local UniversE Simulations \citep[CLUES,][]{Gottloeber2010} and 
MultiDark \citep{Riebe2013}, AIP hosts hundreds of terabytes of file storage and 
a relational database with about 100~TB of carefully curated particle information, 
halo catalogs, and results of semi-analytical galaxy models. These data products 
are available via the 
CosmoSim\footnote{\href{http://www.cosmosim.org}{www.cosmosim.org}} database 
provided by the E-Science group at AIP. However, also observational 
collaborations, e.g., the Multi Unit Spectroscopic 
Explorer\footnote{\href{http://muse-vlt.eu/science}{muse-vlt.eu/science}} 
\citep[MUSE,][]{Bacon2010, Weilbacher2014} collaboration and the Radial Velocity 
Experiment\footnote{\href{http://www.rave-survey.org}{www.rave-survey.org}} 
\citep[RAVE,][]{Steinmetz2006} survey, rely on a CRE for data management and 
processing.

The CRE hardware is integrated into the Almagest cluster at AIP. The whole 
cluster consists of over 50 nodes and about 3~PB of raw disk space. The 
different machines are connected through a high-performance InfiniBand network. 
Throughout the cluster we use the Linux distribution CentOS as operating system. 
Directly allocated for the GREGOR CRE are:
\begin{itemize}
\item[$-$] \textit{One shared compute node acting as login node and to share the data 
    among the collaboration.} Collaborators can log in to this machine 
    using the SSH protocol and copy data to their workstations for further 
    processing and scientific analysis. 
\item[$-$] \textit{Two storage nodes with 80~TB of storage space are reserved for 
    GREGOR data.} Each of the nodes contains 24 hard-disks, which are combined 
    to one logical RAID volume using a 
    ZFS\footnote{\href{http://zfsonlinux.org/}{zfsonlinux.org}}
    file system. Using ZFS's send-receive feature, the disk content of the 
    two nodes is mirrored, and they are physically located in different 
    buildings to minimize the risk of data loss due to catastrophic events.
\item[$-$] \textit{A dedicated compute node, hosting the sTools pipeline, which is 
    directly connected to the archive.} This machine allows users to run 
    their own programs and to visualize GFPI and HiFI (raw) data. This 
    computer is also used for generating and updating the webpages with 
    quick-look data. Remote users can log in by SSH and use Virtual 
    Network Computing (VNC) for a desktop-like environment.
\item[$-$] Additional resources are supplied to host the web application 
    for the data archive, for connecting to the internet, and for backups.
\item[$-$] \textit{Two internal compute nodes with mirrored installations of the 
    sTools data processing pipeline.} One of the nodes with a 
    64-core processor and 256~GB RAM is dedicated to image 
    restoration and hosts the SVN repository of the sTools code including 
    external libraries. The workstations of the researchers on the AIP 
    campus can NFS-mount the respective volumes containing data and software
    libraries so that data processing and analysis is also possible locally. 
\item[$-$] In addition, a copy of the sTools data processing pipeline is
    installed at the German solar telescopes at Observatorio del Teide,
    Iza{\~n}a, Spain. The computer network includes workstations and a 
    dedicated multi-core computer for data processing {and image restoration} 
    on site. In particular, the MFGS-based image selection for HiFI data is 
    carried out immediately after the observations with an easy to use GUI 
    written in IDL as interface to sTools.
\end{itemize}


\subsection{Data Levels}\label{SEC45}

We distinguish three major levels of data products within the GREGOR GFPI and 
HiFI data archive:
\begin{itemize}
\item[$-$] Level~0 refers to raw data acquired with GFPI and HiFI. The data
    are written in a format native to the DaVis software of LaVision, which
    runs both instruments. A short ASCII header declares the basic properties of
    the images (e.g., DaVis version number, image size in pixels, number of
    image buffers, type of image compression, if applicable, etc.), which is
    followed by either compressed or uncompressed binary data blocks. Another
    free-format ASCII header, containing auxiliary information like a time stamp 
    with microsecond accuracy, is placed after the data blocks at the end of the
    file. This time stamp results from the programmable timing unit (PTU),
    which provides external trigger signals for image acquisition. Specific 
    settings of the observing mode and instrument parameters are saved in 
    additional text files for each image sequence or spectral scan. These
    set files include, for example, telescope and AO status as well as a separate
    time stamp for the observing time based on the camera computer's internal 
    clock, which is synchronized with a local GPS receiver at the observatory.
\item[$-$] The huge amount of large-format, high-cadence HiFI data (up to 4~TB 
    per day with the current set-up) requires reducing the data already on 
    site, directly after the observations. This is the standard procedure
    and includes dark and flat-field corrections. In addition, the
    image quality is determined with the MFGS metric, which facilitates
    frame selection. Only the best 100 out of 500 images in a set are kept.
    The calibrated image sequences (level~1) of the two synchronized
    sCMOS cameras are written as FITS files with image extensions. The metadata
    contained in the primary and image headers are partially SOLARNET-compliant
    (level~0.5), which means that they do not contain all mandatory SOLARNET
    keywords, and that they have not used any SOLARNET/FITS standard keywords 
    in a way that is in conflict with their definitions. HiFI level~0
    data are typically deleted, once the frame-selected and calibrated level~1
    data are safely stored in the GREGOR archive. HiFI level~0 images are only kept
    when the seeing condition were excellent, when fast or transient events were
    captured, or when special observing programs were carried out 
    \citep[see][]{Denker2018b}. The processing time for restoring a single
    HiFI image (level~2) from a set of 100 images with KISIP takes several
    tens of minutes on the 64-core compute node.
\item[$-$] Level~1 GFPI data are typically created at AIP after the observing 
    campaign, which reflects the high complexity of data from imaging 
    spectroscopy. In most cases, the multiple images per wavelength point 
    are simply destretched and co-added with no image restoration. However, 
    all other calibration steps (e.g., alignment of narrow- and broad-band
    images, blueshift and prefilter-curve corrections) are carried out so that
    scientific exploitation of level~1 data is possible. After preprocessing
    level~1 data, i.e., after dark and flat-field corrections and determining the
    alignment of narrow- and broad-band images, a copy of narrow- and broad-band
    images is saved, which serves as the starting point for image restoration 
    (level~2), thus avoiding preprocessing level~0 data twice. Only the 
    best spectral scans are then chosen for level~2 processing, i.e., image
    restoration with MOMFBD or speckle deconvolution. The processing time of 
    a single scan with MOMFBD takes several hours on the 64-core compute node.
\item[$-$] Level~1 data are the starting point for creating quick-look data
    products such as time-lapse movies, Doppler velocity maps, and overview
    graphics for seeing conditions and observing parameters.
\item[$-$] Level~2 data are restored data using MOMFBD and KISIP (see 
    Sect.~\ref{SEC3}). These data processing steps are only included on demand,
    considering the significant amount on computational resources. The best
    image restoration scheme is chosen by the researcher working with the data,
    and it is not unusual to select different ways to restore images or spectral
    scans depending on user preferences or a specific science case. Once
    the spectral scans are restored, other calibration steps still need to be
    applied such as blueshift and prefilter-curve corrections, before physical
    parameters such as Doppler velocities or other spectral line properties 
    can be determined.
\end{itemize}

Level~1 HiFI data (and occasionally level~0) and level~0 GFPI data are 
transferred from the GREGOR telescope to AIP by regular 2.5-inch external 
hard-disk drives with 2\,--\,4~TB storage capacity. Smaller data sets are often 
transferred over the internet. However, the physical transfer using hard-disks 
is preferred over copying over the internet due to limited bandwidth and network 
reliability at the observatory site. Data from all GREGOR instruments can be 
stored on site on a 100~TB storage array for up to three months. Keeping at least
two copies of the data at different locations during the data transfer to the
GREGOR archive mitigates against potential data loss. Data integrity during the
transfer is assured by monitoring the transfer logs and verifying that all files
with correct sizes were transferred. The total amount of data, which was acquired 
with GFPI and HiFI, as well as with the now obsolete facility cameras of BIC, is
summarized in the last row of Table~\ref{TAB01}. The other entries in the rows 
for 2014\,--\,2016 refer to the number of scientific data sets, which were 
obtained in various observing campaigns. The data volume refers to the sum
over all data levels. GFPI level~1 data is roughly twice the size than level~0
data. Furthermore, the bulk of the data volume arises from level~0 data 
for BIC and from level~1 data for HiFI.

Finally, 70 users are currently registered in the CRE, who are mainly 
from the GREGOR consortium. However, increasingly external
users (at the moment ten) register with the CRE, most of them 
participated in (coordinated) observing campaigns or in the SOLARNET Access  
Program promoting observing campaigns with
Europe's telescopes and instruments for high-resolution solar physics. In the
meantime, routine observation started at GREGOR in 2016, and we already see an 
influx of new international collaborators, who will certainly broaden the user 
base of the GREGOR CRE.


\subsection{Use Case: GREGOR Early Science Phase}\label{SEC46}

The GREGOR early science phase took place in 2014 and 2015, where members of the 
GREGOR consortium collaboratively carried out observing 
campaigns, i.e., in 2014 with the individual instruments and in 2015 with 
multi-instrument set-ups. Notably, a 50-day observing campaign with GFPI and BIC 
(the predecessor of HiFI) was carried out in July and August 2014. In a joint 
effort, scientists from all involved institutes submitted observing proposals, 
which were evaluated and condensed into a list of top-priority solar targets and 
feasible observing modes and strategies. The observations were carried out by 
experienced observers of all institutes together with novice-observers (not 
necessary novice-scientists) to strengthen their observing skills and to 
familiarize them with the new instruments. The data, which were acquired with 
many different set-ups, were used to develop, test, and improve the sTools data 
processing pipeline. Finally, the level~1 data were stored in the GREGOR GFPI 
and HiFI data archive so that all scientists were able to access and analyze the 
data. Collaborations on specific studies, with the aim of publishing them in 
scientific journals, were coordinated using the blog facilities of the Daiquiri 
framework. This includes also the organization of GREGOR science meetings at 
AIP. The News Blog of the GREGOR CRE allowed us to publish poster presentations 
given at solar physics meetings (e.g., the SOLARNET~IV Meeting in Lanzarote in 
2016) and newly appearing journal articles and conference proceedings to a 
broader section of the solar physics community and the interested public. First
results of the GREGOR early science phase were published in 2016 in the journals
\textit{Astronomy \& Astrophysics} (Vol.~596) and \textit{Astronomische 
Nachrichten} (Vol.~337/10), and more up-to-date GFPI and HiFI science 
publication are referenced in Sect.~\ref{SEC22}.


\subsection{Future Plans}\label{SEC47}

In the future, we plan to expand the data archive and data access infrastructure 
considerably. With a release of the data to the general solar community, new 
means of data access will be necessary, beyond users actively utilizing the
CRE to interact with like-minded researchers. As before, the latter type of
access will be managed via registration and subsequent confirmation by the
collaboration. While collaborators will still be able to retrieve data 
through SSH connections, this is not suitable for public access by the general
scientific community. 

Therefore, we will extend the GREGOR archive towards a public data portal, which will offer
a full search on all files in the archive, unless they are
still embargoed, and downloads using the HTTP protocol, either through the browser or using
command line tools. The selection of files will be based on the metadata of level~1 and~2 data of the GFPI and HiFI instruments.
The archive will also allow for SQL queries to create custom result sets based on any scientific criteria.
Standards defined and adopted by IVOA like UWS and
Table Access Protocol (TAP) will expedite accessing the data through clients using
interoperable Application Programming Interfaces (APIs). The results of the
queries of a user and the queries themselves will be stored in a personal
database to be retrieved at the user's convenience without repeating the query
to the whole GREGOR database.

\begin{table}[t]
\caption{Number of data sets and stored data volume for GREGOR's instruments.}
\begin{center}
\begin{tabular}{cccc}
\hline\hline
     & \textbf{GFPI} & \textbf{BIC} & \textbf{HiFI}\rule[-6pt]{0pt}{16pt} \\
\hline
2014 &     48        &     37       &        \\
2015 &     69        &     42       &        \\
2016 &     41        &  \phn 4      &     46 \\
2017 &     35        &              &     17 \\
\hline
     &   10.0~TB     &   13.6~TB    & 20.3~TB\rule[-6pt]{0pt}{16pt}\\
\hline
\end{tabular}
\end{center}
\label{TAB01}
\end{table}

%
%

\section{Conclusions and Prospects}\label{SEC5}

High-resolution solar observations are confronted with short time-scales in both 
the Sun's and the Earth's atmospheres, whereas the first is related to evolution 
and dynamics of solar features, the latter imposes strong boundary conditions 
for image restoration. The underlying assumption is that the observed object is 
not changing. Thus, a contiguous data set has to be recorded within a few 
seconds. This interval becomes even shorter when observing faster features like 
in the chromosphere or with larger telescopes offering higher spatial resolution 
and consequently smaller ``diffraction-limited'' pixels. These time-scales 
affect instrument design, observing modes and strategies, and in the end also 
data management and archiving.

Providing high-resolution data to the solar community is an ongoing process, 
i.e., the GREGOR GFPI and HiFI archive and the sTools data processing pipeline 
are not static. At present, we are working on the polarimetric calibration 
routines, and they will be added to the sTools package once they are tested. The 
CRE implemented for the GREGOR telescope takes this into account and also 
contains provisions for future data access beyond the 
consortium and collaboration. Currently, our efforts are focused on 
characterizing image quality to identify the best data sets of HiFI images and 
GFPI spectral scans, which are both affected by the varying seeing conditions
\citep[see][]{Denker2018b}. 
The algorithms and routines (e.g., with the MFGS method)  developed for this 
purpose additionally allow us to monitor long-term trends in data quality as 
well as to establish a database of seeing conditions at Observatorio del Teide.

Synoptic solar images, magnetograms, and Doppler maps with a typical spatial 
resolution of one second of arc serve very successfully as input for feature 
identification and pattern recognition algorithms. In particular, in the context 
of space weather research and forecasting tools were developed to enable easy 
access to such data products, e.g., the Heliophysics Event Knowledgebase 
\citep[HEK,][]{Hurlburt2012}. 

Even though the focus shifts from applications to 
more fundamental physics, a knowledgebase for high-resolution data (spatial 
resolution below 0.1\arcsec) is potentially very beneficial, bringing order into 
the plethora of small-scale features observed in the quiet-Sun (e.g., G-band 
bright points, filigree, blinkers, etc.) and in active regions (e.g., umbral 
dots, penumbral grains, Ellerman bombs, micro-flares, etc.). 
Morphological characteristics, photometric properties, and spectral/polarimetric features 
provide a wide parameter range, which can be stored in relational databases.
A summary of image processing techniques and various ways of
performing feature tracking is given in \citet{Aschwanden2010},
and \citet{Turmon2010} demonstrated the capability of multidimensional
feature identification. Database research can reveal relationships among solar 
small-scale features such as those of the photospheric network, which would otherwise 
be missed in case studies, and can track changes with the 
solar activity cycle, when the contents of the database grows over time 
\citep[e.g.,][]{McIntosh2014, Muller1994, Jin2011, Roudier2017}.

The GREGOR GFPI and HiFI archive and in the future associated relational 
databases are an attractive starting point for data mining and machine learning 
applications. In general, the huge amount of astronomical and astrophysical data 
has stimulated interest effectively exploring them \citep[see][for current 
research in this field]{Ball2010, Ivezic2014}. Machine learning algorithms gain 
knowledge from experience. A training data set teaches the underlying models to 
the machine, and new data sets can then be classified with the results from the 
initial training set. Furthermore, machine learning can be used in time-series 
and wavelet analysis. Data mining often employs machine learning techniques when 
data sets become overwhelmingly large, extracting useful information from raw 
data and detecting new relations or anomalies. Furthermore, data mining and 
machine learning help validating model assumptions and ensure consistency.
The performance of machine learning models is mostly influenced by
the amount and quality of the training data sets. A central repository with 
immediate access to calibrated high-resolution solar data, such as ours, speeds 
up the process of training neural networks or data mining.

As an example, DeepVel \citep{AsensioRamos2017} is a deep learning neuronal 
network, estimating horizontal velocities at three different atmospheric heights 
from time-series of high-resolution images. The training data were in this case 
numerical simulations. The advantage of DeepVel is not only its extremely fast 
execution, as compared to other optical flow techniques, but also its the 
ability to infer velocity fields from subphotospheric layers. Machine learning 
will become increasingly helpful in image restoration and spectral inversions -- 
both applications requiring significant computational resources.

%
%

\acknowledgments
\renewcommand{\baselinestretch}{0.9}
\noindent
\parbox{\columnwidth}{\small\hspace*{5pt}
The 1.5-meter GREGOR solar telescope was built by a German 
consortium under the leadership of the Kiepenheuer Institute for Solar Physics 
in Freiburg with the Leibniz Institute for Astrophysics Potsdam, the Institute 
for Astrophysics G\"ottingen, and the Max Planck Institute for Solar System 
Research in G\"ottingen as partners, and with contributions by the Instituto de 
Astrof\'{\i}sica de Canarias and the Astronomical Institute of the Academy of 
Sciences of the Czech Republic. We thank Dr.\ Michiel van Noort for his
help in implementing the MOMFBD code at AIP. SJGM acknowledges support of 
project VEGA 2/0004/16 and is grateful for financial support from the 
Leibniz Graduate School for Quantitative Spectroscopy in Astrophysics, a joint 
project of the Leibniz Institute for Astrophysics Potsdam and the Institute of 
Physics and Astronomy of the University of Potsdam. CD and REL were supported by 
grant DE 787/3-1 of the German Science Foundation (DFG). This study is supported 
by the European Commission's FP7 Capacities Program under Grant Agreement number 
312495. The AIP Almagest cluster and its CREs were partially funded by an  
European Regional Development Fund (ERDF) grant in 2012. Development of VO 
interfaces and facilities have been supported by the German Federal Ministry of 
Education and Research (BMBF) Collaborative Research Program for the German 
Astrophysical Virtual Observatory (GAVO). Development of Daiquiri has been 
partially supported by a BMBF grant ``Survey-Competence''. The Center of
Excellence in Space Sciences India is funded by the Ministry of Human Resource 
Development, Government of India.}\bigskip
\renewcommand{\baselinestretch}{1.0}\normalsize

\facilities{GREGOR solar telescope (GFPI, HiFI)}\\

\software{%
    DeepVel \citep{AsensioRamos2017},
    KISIP \citep{Woeger2008a}, 
    MOMFBD \citep{Loefdahl2002, vanNoort2005}, 
    MPFIT \citep{Markwardt2009},
    SolarSoft \citep{Bentley1998, Freeland1998}, and
    sTools \citep{Kuckein2017a}}

%
%

\end{document}